\documentclass[11pt, oneside]{article}
\usepackage{jheppub}
\usepackage{amsmath}
\usepackage{amssymb}
\usepackage{amsfonts}
\usepackage{amsthm}
\usepackage{amsbsy}
\usepackage{array}
\usepackage{mathtools}
\usepackage[usenames,dvipsnames]{xcolor}
\usepackage{caption}
\usepackage{subcaption}
\usepackage{dsdshorthand}
\usepackage{graphicx}
\usepackage[vcentermath]{youngtab}
\usepackage{tikz}
\usepackage{booktabs,multirow}
\usepackage{listings}
\usepackage{xcolor}
\usepackage{tikz}
\usepackage{subcaption}
\usetikzlibrary{calc}

\definecolor{darkblue}{rgb}{0,0,0.5}
\definecolor{darkgreen}{rgb}{0,0.3,0}
\definecolor{darkpink}{rgb}{0.6,0,0.4}
\definecolor{graygreen}{rgb}{0.3,0.5,0.3}
\definecolor{grayblue}{rgb}{0.2,0.2,0.6}
\definecolor{grayred}{rgb}{0.5,0.2,0.2}

\newcommand*\link[1]{\hspace*{0em plus 1fill}\makebox{#1}}

\makeatletter
\def\@fpheader{\ }
\makeatother

\title{Bootstrapping currents and stress tensors in 3d CFTs}
\author{Vasiliy Dommes$^1$, Rajeev S.~Erramilli$^{2,3}$, Colum Flynn$^4$ Alexandre Homrich$^{1,5}$, Petr Kravchuk$^4$, David Poland$^6$, David Simmons-Duffin$^1$,
George Yuzhe Zhou$^1$}
\affiliation{$^1$Walter Burke Institute for Theoretical Physics, Caltech, Pasadena, California 91125, USA}
\affiliation{$^2$Institut des Hautes \'Etudes Scientifiques, 91440 Bures-sur-Yvette, France}
\affiliation{$^3$Leinweber Institute for Theoretical Physics, Randall Laboratory of Physics, University of Michigan, Ann Arbor, MI 48109, USA}
\affiliation{$^4$Department of Mathematics, King’s College London, Strand, London, WC2R 2LS, UK}
\affiliation{$^5$Kavli Institute for Theoretical Physics, University of California, Santa Barbara, CA 93106, USA}
\affiliation{$^6$Department of Physics, Yale University, 217 Prospect St, New Haven, CT 06520, USA}
\emailAdd{vdommes@caltech.edu}
\emailAdd{rserram@umich.edu}
\emailAdd{colum.flynn@kcl.ac.uk}
\emailAdd{ahomrich@caltech.edu}
\emailAdd{petr.kravchuk@kcl.ac.uk}
\emailAdd{david.poland@yale.edu}
\emailAdd{dsd@caltech.edu}

\date{}
\abstract{We perform a numerical conformal bootstrap study of the mixed system of correlation functions involving a spin-1 $\mathrm{U}(1)$ current $J$ and the stress-energy tensor $T$ in parity-preserving 3d CFTs. We find universal bounds on the stress-tensor two-point function $c_T$, which numerically reproduce the conformal collider bounds on $\<JJT\>$ and $\<TTT\>$, as well as bounds on the leading parity-even and parity-odd scalar operator dimensions. Under mild assumptions, we determine the values of the $\<JJT\>$ and $\<TTT\>$ three-point functions in the $\mathrm{O}(2)$ vector model. The $\<TTT\>$ result is new and $\<JJT\>$ is consistent with a previous result in the literature.}
\preprint{CALT-TH 2026-031\\
\link{LITP-26-18}}

\begin{document}

\maketitle
\pagenumbering{roman}
\setcounter{page}{2}
\newpage
\pagenumbering{arabic}
\setcounter{page}{1}

\setlength{\parskip}{1pt}

\section{Introduction}

Numerical conformal bootstrap is a powerful method for constraining the properties of conformal field theories (see \cite{Poland:2018epd,Rychkov:2023wsd} for recent reviews). It has been especially successful in $d=3$ spacetime dimensions, where it provides high-precision determinations of critical exponents in conformal field theories such as 3d Ising, $\mathrm{O}(N)$, Gross-Neveu-Yukava, and other models~\cite{El-Showk:2012cjh,El-Showk:2014dwa,Simmons-Duffin:2016wlq,Kos:2016ysd,Kos:2014bka,Kos:2013tga,Kos:2015mba,Chang:2024whx,Erramilli:2022kgp,Iliesiu:2015qra,Iliesiu:2017nrv,Chester:2019ifh,Chester:2020iyt}. Early works on conformal bootstrap focused on correlation functions of scalar operators. More recently, correlation functions with spin have been incorporated~\cite{Iliesiu:2015qra,Iliesiu:2017nrv,Dymarsky:2017yzx,Dymarsky:2017xzb,Karateev:2019pvw,Karateev:2025sjw,Reehorst:2019pzi,He:2023ewx,Bartlett-Tisdall:2024mbx,Mitchell:2024hix,Bartlett-Tisdall:2023ghh,DiPietro:2025ozw,Erramilli:2026ykj}, leading both to results for new theories and to significant improvements of existing constraints. This progress has been supported by significant improvements to and the development of new numerical codes and methods~\cite{Erramilli:2019njx,Erramilli:2020rlr,Chang:2025mwt,Landry:2019qug,Liu:2023elz,Reehorst:2021ykw}.

In this paper we continue the exploration of the ``spinning'' numerical conformal bootstrap in $d=3$ by studying the mixed system of correlation functions for a $\mathrm{U}(1)$ spin-1 current $J^\mu$ and the stress-energy tensor $T^{\mu\nu}$, assuming space parity invariance. Similarly to earlier works on the pure stress-tensor~\cite{Dymarsky:2017yzx} and the pure current~\cite{Dymarsky:2017xzb} systems, the bounds we derive using this system are universal and apply to any local parity-invariant unitary conformal field theory with $\mathrm{U}(1)$ global symmetry and are therefore useful for constraining the space of CFTs.

For instance, we obtain unconditional bounds on the stress-tensor two-point function coefficient $c_T$ without assuming the existence of scalar operators of any particular scaling dimension (figure~\ref{fig:centralCharge}). When computed as a function of the ``shape'' parameters $\g_{JJT}$ and $\g_{TTT}$ of $\<JJT\>$ and $\<TTT\>$ three-point functions, our $c_T$ bounds reproduce the conformal collider bounds~\cite{Hofman:2008ar,Hofman:2016awc} on $\g_{JJT}$ and $\g_{TTT}$ (figure~\ref{fig:conformalColliderBound}). We also find unconditional bounds on the scaling dimensions of the lightest parity-even and parity-odd scalar operators which are somewhat stronger than those from either the pure $J$ or pure $T$ system~\cite{Dymarsky:2017xzb,Dymarsky:2017yzx} (figure~\ref{fig:scalarGaps}).

Another motivation for studying the $JT$ system of crossing equations is that it is a subsystem of the natural extension of the previous scalar correlator studies of the $\mathrm{O}(2)$ model~\cite{Kos:2013tga,Kos:2015mba,Kos:2016ysd,Chester:2019ifh}. We expect that considering the combination of $JT$ and the systems of~\cite{Kos:2013tga,Kos:2015mba,Kos:2016ysd,Chester:2019ifh} will give significant improvements in precision of the bounds on the $\mathrm{O}(2)$ model, similarly to what was demonstrated for 3d Ising CFT in~\cite{Chang:2024whx}. Furthermore, by replacing the abelian current $J$ with its non-abelian versions one can explore the landscape of theories with non-abelian symmetry, or add additional operators to focus on $\mathrm{O}(N)$ or other models. We therefore view the current work also as a preparation for these future studies.

In fact, the $JT$ system itself is already sensitive to new observables in $\mathrm{O}(2)$ model. In particular, under mild spectral assumptions detailed in section~\ref{sec:O2}, we show that the parameters $\g_{JJT}$ and $\g_{TTT}$ in the $\mathrm{O}(2)$ model satisfy
\begin{align}
	\gamma_{JJT}=-0.0776(40),\quad \gamma_{TTT}=-0.0790(29),
\end{align}
see section~\ref{sec:WardIds} for the precise definitions. This provides the first nonperturbative determination of $\g_{TTT}$ in the $\mathrm{O}(2)$ model. This result can be somewhat improved using the determination of $\g_{JJT}$ from~\cite{Reehorst:2019pzi}, see section~\ref{sec:O2} for details.

This paper is organized as follows. We describe the details of our setup in section~\ref{sec:setup}. The results are presented in section~\ref{sec:results}. We conclude with a discussion in section~\ref{sec:discussion}. The appendices contain some technical details of the setup.

\section{Setup}
\label{sec:setup}

\subsection{Choice of correlators and symmetry enhancement}
\label{sec:symmetry}

In this work, we study the mixed system of four-point functions of a $\mathrm{U}(1)$ current $J$ and the stress tensor $T$. In principle, it should include the correlation functions
\begin{align}\label{eq:setnaive}
	\<JJJJ\>,\quad \<TJJJ\>,\quad \<TTJJ\>,\quad \<TTTJ\>,\quad \<TTTT\>.
\end{align}
We claim that a large class of bounds for this system can be just as well computed in the smaller system containing only the correlators with an even number of $J$ insertions:
\begin{align}\label{eq:setsmart}
	\<JJJJ\>,\quad \<TTJJ\>,\quad \<TTTT\>,
\end{align}
and assuming symmetry under $\Z_2$ charge conjugation that sends $J\to -J$. This is an example of symmetry enhancement of numerical bootstrap bounds~\cite{Poland:2011ey, Li:2020bnb, Kravchuk:2021akc}.

To see that this this the case, we follow the explanation of symmetry enhancement given in~\cite{Kravchuk:2021akc}. Let $p$ denote the full OPE density of conformal block expansions of the above correlation functions; $p$ should be seen as a matrix-valued positive-semidefinite measure on the set of all possible contributions to the conformal block expansions. For instance, parity-even spin-4 contributions to the $\<TTTJ\>$ four-point function in the $JT\to TT$ OPE channel are parametrized by their scaling dimension $\De$, and $p$ provides a matrix-valued measure $dp_{ab,JT\to TT,\text{spin-4}}(\Delta)$. The indices $a,b$ label the OPE coefficients corresponding to the different tensor structures appearing in $JT$ and $TT$ OPEs, respectively.

Formally, any four-point function in~\eqref{eq:setnaive} can be written as
\begin{align}
	\<\cO_1\cO_2\cO_3\cO_4\> = \int_{X_{12|34}} dp_{ab} G^{ab}_{1234},
\end{align}
where $X_{12|34}$ is the set of all unitary contributions to $\cO_1\x \cO_2\to \cO_3\x\cO_4$ OPE channel, and $G^{ab}_{1234}$ is the corresponding conformal block.

The key observation is that if $p_1$ and $p_2$ are two measures of the above kind, that solve all the crossing equations in the system~\eqref{eq:setnaive}, then $tp_1+(1-t)p_2$ for $t\in [0,1]$ is also a measure of the same kind, properly normalized, and solving the same crossing equations. This allows us to generate more symmetric solutions to crossing by averaging the less symmetric solutions to crossing over a symmetry group.

In particular, let $p$ be a solution to crossing equations of~\eqref{eq:setnaive} and let $p'$ be obtained from $p$ by replacing $J\to -J$. It is easy to see that $p'$ is also a positive-semidefinite solution. Then $p_s=\half p+\half p'$ is a new solution which is now invariant under $J\to -J$. In this way we obtain a solution of crossing equations for the charge conjugation-invariant system~\eqref{eq:setsmart} from a solution of more general system~\eqref{eq:setnaive}. Conversely, any solution of crossing for the charge conjugation-invariant system~\eqref{eq:setsmart} is, trivially, a solution of the crossing equations for the more general system~\eqref{eq:setnaive} by simply forgetting the symmetry.

Therefore, the two systems are equivalent, provided any extra assumptions that we use in our bounds are also mapped by the averaging procedure. For instance, if we would like to study a gap maximization problem for parity-even spin-4 operators for~\eqref{eq:setnaive}, this is equivalent to studying the gap maximization problem for~\eqref{eq:setsmart}, where the same gap is imposed both in the charge conjugation-even and charge conjugation-odd versions of the parity-even spin-4 sectors. This works because the objective---the dimension gap---is invariant under the enhanced symmetry transformation. Objectives for~\eqref{eq:setnaive} that aren't invariant might not have equivalent versions for~\eqref{eq:setsmart}.

In what follows, we will focus on the system~\eqref{eq:setsmart} with charge conjugation symmetry and then specify explicitly which of our bounds also apply to the full~\eqref{eq:setnaive} system.

\subsection{Four-point functions and crossing equations}

Next, we must identify a basis of four-point tensor structures determining the correlators in the $JT$ system~\eqref{eq:setsmart}.
The computation is a straightforward generalization of the analysis described in
\cite{Kravchuk:2016qvl,Dymarsky:2017xzb,Dymarsky:2017yzx,Chang:2024whx}. Here we merely summarize the results.

To describe the structures in the correlator $\langle \mathcal{O}_1 \mathcal{O}_2 \mathcal{O}_3 \mathcal{O}_4 \rangle$ we will use the symmetric $q$-basis $\langle q_1 q_2 q_3 q_4\rangle^\pm$ of \cite{Kravchuk:2016qvl}, to which we refer for details.
Here $q_i \in \{-\ell_i, \dots, \ell_i\}$, where $\ell_i$ is the spin of the operator $O_i$. The index $\pm$ denotes the signature of the structure under $q_i\leftrightarrow -q_i$.

The $q$-basis elements are constructed to realize the appropriate conformal-invariance constraints on tensor structures.  They also enforce the invariance of the correlator under kinematic permutations, i.e.~those that leave $z,\bar{z}$ fixed. In our $JT$ system, these are:
\begin{align}
	\Pi_{TTTT} & = \Pi_{JJJJ} = \{\text{id}, (12)(34), (13)(24),(14)(23)\}, \\
	\Pi_{JJTT} & = \{\text{id}, (12)(34)\},
\end{align}
with analogous expressions for $JTTJ$ and $JTJT$. Finally, parity invariance of the correlator constrains $\sum_i q_i \in 2 \mathbb{Z}$.

The correlation function can thus be written as
\begin{equation}
	\langle \mathcal{O}_1 \mathcal{O}_2 \mathcal{O}_3 \mathcal{O}_4 \rangle = \sum_{\underset{\sum_i q_i \in 2\mathbb{Z}}{\{q_i\}/{\Pi, \pm}}} \langle q_1 q_2 q_3 q_4\rangle^\pm g_{\mathcal{O}_1\mathcal{O}_2\mathcal{O}_3\mathcal{O}_4}^{{\langle q_1 q_2 q_3 q_4 \rangle}^\pm}(z,\bar{z}). \label{eq:4pt-structures}
\end{equation}
where $g_{\mathcal{O}_1\mathcal{O}_2\mathcal{O}_3\mathcal{O}_4}^{{\langle q_1 q_2 q_3 q_4 \rangle}^\pm}(z,\bar{z}) = \pm g_{\mathcal{O}_1\mathcal{O}_2\mathcal{O}_3\mathcal{O}_4}^{{\langle q_1 q_2 q_3 q_4 \rangle}^\pm}(\bar{z},z)$.
The total number of tensor structures appearing in (\ref{eq:4pt-structures}) is summarized in table \ref{tb:structures}.
The functions $g_{\mathcal{O}_1\mathcal{O}_2\mathcal{O}_3\mathcal{O}_4}^{{\langle q_1 q_2 q_3 q_4 \rangle}^\pm}(z,\bar{z})$ are further constrained by the conservation equations of $J$ and $T$,
\begin{equation}
	\langle \partial_\mu J^{\mu}\dots\rangle = \langle \partial_\mu T^{\mu\nu}\dots\rangle = 0,
\end{equation}
which we now describe.

\begin{table}[t]
	\centering
	\renewcommand{\arraystretch}{1.15}
	\begin{tabular}{l||c|c|c||c|c|c}
		\hline
		 & conformal & parity & permutation & bulk & line & point \\
		\hline
		$\langle TTTT \rangle$
		 & 625       & 313    & 97          & 5    & 9    & 8     \\

		$\langle JJJJ \rangle$
		 & 81        & 41     & 17          & 5    & 2    & 0     \\

		$\langle TTJJ \rangle$
		 & 225       & 113    & 64          & 6    & 5    & 3     \\
		\hline
	\end{tabular}
	\caption{
		\label{tb:structures} The number of independent functional degrees of freedom contributing to the correlators in our $JT$ system as conformal symmetry, parity invariance, permutation invariance, and $J,T$ conservation are imposed.}
\end{table}

As is now standard \cite{Kravchuk:2016qvl,Dymarsky:2017xzb,Dymarsky:2017yzx,Chang:2024whx},
we interpret the conservation equations as a set of PDEs
for the coefficient functions $g_{\mathcal{O}_1\mathcal{O}_2\mathcal{O}_3\mathcal{O}_4}^{{\langle q_1 q_2 q_3 q_4 \rangle}^\pm}$.
These PDEs determine all coefficient functions, and hence the full correlator
$\langle \mathcal{O}_1\mathcal{O}_2\mathcal{O}_3\mathcal{O}_4 \rangle$, once a minimal set of ``bulk'', ``line'', and ``point''
variables is fixed.
The bulk variables are unconstrained on the full $(z,\bar{z})$ plane and act as source terms for the conservation equations,
while the line and point degrees of freedom serve as initial data on the line $t \equiv ({z-\bar{z}})/{2i} = 0$.
This initial data is further constrained\footnote{In principle, one must also demand the regularity of the correlator around the $t=0$ line.
	This might enforce that certain $t$-derivatives of the $g_{\mathcal{O}_1\mathcal{O}_2\mathcal{O}_3\mathcal{O}_4}^{{\langle q_1 q_2 q_3 q_4 \rangle}^\pm}$
	variables to vanish, see \cite{Kravchuk:2016qvl} for details. Similarly to~\cite{Dymarsky:2017yzx}, such constraints are automatically satisfied in our setup by solutions to conservation equations.} by subsets of the
conservation equations that close on the line $t=0$.
Here, the line degrees of freedom act as source terms, and must be specified on the full $t=0$ line, while point degrees of freedom need
only to be specified at the crossing-symmetric point $z = \bar{z} = 1/2$ and serve as boundary data.
The total numbers of independent bulk, line and point variables in each correlator are provided in table \ref{tb:structures}.

The crossing symmetry constraints must\footnote{Indeed, it is not only sufficient but also necessary to impose crossing symmetry only on
	independent functional variables $g_{\mathcal{O}_1\mathcal{O}_2\mathcal{O}_3\mathcal{O}_4}^{{\langle q_1 q_2 q_3 q_4 \rangle}^\pm}$.
	The reason is that spurious crossing equations lead to nearly-flat directions in the SDP, causing numerical instabilities
	that can render bounds invalid \cite{Chang:2024whx}.}
be imposed on this minimal set of $g_{\mathcal{O}_1\mathcal{O}_2\mathcal{O}_3\mathcal{O}_4}^{{\langle q_1 q_2 q_3 q_4 \rangle}^\pm}$ variables.
In total, our $JT$ system requires 28 crossing equations on bulk variables, 26  crossing equations on line variables, and 8 crossing equations
on point variables. The complete list of crossing equations is provided in appendix \ref{ap:crossing}.
Let us point out that our analysis implies two additional line constraints (and two fewer point constraints) in the $JJJJ$ correlator
relative to the analysis of \cite{He:2023ewx}, in agreement with \cite{Bartlett-Tisdall:2023ghh}.
In particular, note that any particular choice of
candidate bulk, line and point free variables can be verified by expanding the conservation
equations around the crossing-symmetric point and checking order by order that all and only the remaining
variables are fixed by the conservation equations. We have performed this check in a notebook attached to this submission.

\subsection{Three-point structures}
We now discuss the classification of the three-point structures involving $J$, $T$, and exchanged operators $\cO$. The procedure is the same
as in \cite{Kravchuk:2016qvl,Chang:2024whx}, by computing $q$-basis structures satisfying conformal,
parity, and permutation invariance, and conservation of $T$ and $J$.

As noted in \cite{Chang:2024whx}, it is important that for any given exchange operator allowed by the unitarity bounds, the chosen three-point structures are linearly independent. When naively solving the $\Delta$-dependent conservation equations that implement conservation of $T$ and $J$, the resulting $\Delta$-dependent structures, although linearly-independent for generic $\Delta$, may have linear dependencies at special values of $\Delta$. We found that a good strategy to avoid this problem is as follows. The conservation equations are polynomial in $\De$, and so the solutions can be chosen to be polynomial as well. We then search for a solution with the minimal degree in $\Delta$ by writing down low-degree polynomial ans\"atze for the three-point structures and trying to solve for the unknown coefficients. This tends to produce a solution free from the above problem. Another possibility is to use Smith normal form of the conservation equations; however, the approach described above has the benefit of finding the lowest-degree solution, which is useful for runtime performance.

The summary of three-point tensor structures relevant for the $T \times T$, $J \times J$, and $J \times T$
OPEs can be found in appendix \ref{ThreePointStructures}. We now turn our attention to the cases where $T$ or $J$ appear in the OPE, in which conformal Ward identities for the stress tensor impose additional constraints on the three-point functions.

\subsection{Ward identities}
\label{sec:WardIds}
The three-point structures where stress-tensor Ward identities impose additional constraints are $\<TTT\>$ and $\<JJT\>$. These have been studied before in \cite{Osborn:1993cr,Dymarsky:2017xzb,Dymarsky:2017yzx}.

First, let us define a normalized stress tensor and current:
\begin{equation}
	\hat{T}\equiv\frac{T}{\sqrt{c_T}}, \quad \hat{J}\equiv\frac{J}{\sqrt{c_J}},
\end{equation}
where $c_T$ and $c_J$ are the coefficients appearing in the two-point functions
\begin{align}
	\<T^{\mu\nu}(x_1) T^{\r\s}(x_2)\> & =c_T \half \frac{I^{\mu\r}(x_{12})I^{\nu\s}(x_{12})+I^{\mu\s}(x_{12})I^{\nu\r}(x_{12})-\text{traces}}{x_{12}^6}
	,                                                                                                                                                   \\
	\<J^\mu(x_1)J^\nu(x_2)\>          & = c_J \frac{I^{\mu\nu}(x_{12})}{x_{12}^4},
\end{align}
and $I^{\mu\nu}(x)=\de^{\mu\nu}-2x^\mu x^\nu/x^2$. This normalization is standard in the numerical bootstrap and is necessary to ensure that the identity operator contributes with coefficient $1$ to the correlators in~\eqref{eq:setsmart}. Furthermore, the same normalization is also necessary when $T$ or $J$ is the exchanged operator.

The two three-point structures can be decomposed into a sum of the free complex scalar and free Dirac fermion structures:
\begin{equation}
	\begin{aligned}
		\<\hat{T}\hat{T}\hat{T}\> & =\lambda_F\<\hat{T}\hat{T}\hat{T}\>_F+\lambda_B\<\hat{T}\hat{T}\hat{T}\>_B,   \\
		\<\hat{J}\hat{J}\hat{T}\> & =\lambda'_F\<\hat{J}\hat{J}\hat{T}\>_F+\lambda'_B\<\hat{J}\hat{J}\hat{T}\>_B,
	\end{aligned}
\end{equation}
where the structures in the right-hand side are normalized using the $c_J$ and $c_T$ values in the respective theories.
The stress-tensor Ward identities impose the following constraint amongst the OPE coefficients:
\begin{equation}
	\lambda_B+\lambda_F=\lambda'_B+\lambda'_F=\sqrt{\frac{2c_B}{c_T}}
\end{equation}
where
\begin{equation}
	c_B=c_F=\frac{3}{2(4\pi)^2}
\end{equation}
is the central charge of a free real\footnote{Note the factor of 2!} scalar. Note that although $\l'_B$ and $\l'_F$ parameterize the $\<JJT\>$ three point function, whose Ward identity contains $c_J$, the factors of $c_J$ drop out because of the normalization of the operators.

We end up with three undetermined parameters, which we take to be $c_T, x, $ and $y$, and are related to $\l_B, \l_F, \l_B',$ and $\l_F'$ by
\begin{align}
	\l_F  & =\sqrt{\frac{2c_B}{c_T}}x, \quad
	\l_B=\sqrt{\frac{2c_B}{c_T}}(1-x),       \\
	\l_F' & =\sqrt{\frac{2c_B}{c_T}}y, \quad
	\l_B'=\sqrt{\frac{2c_B}{c_T}}(1-y).
\end{align}
Here, $x$ and $y$ are related to $\gamma_{TTT}$ and $\gamma_{JJT}$ by
\begin{equation}\label{eq:xy}
	\begin{aligned}
		x & =\frac{1+12\gamma_{TTT}}{2}, \\
		y & =\frac{1+12\gamma_{JJT}}{2}
	\end{aligned}
\end{equation}
where the conformal collider bounds in terms of $x$ and $y$ are given by
\begin{equation}\label{eq:HMbounds}
	0\leq x,y \leq 1.
\end{equation}

In our bootstrap setup, we are perfectly allowed to set $x$ and $y$ to values outside these bounds. However, as we will see, the bootstrap is still able to approximately
reproduce these bounds.

Finally, note that in our setup we do not have access to $c_J$, the $\<JJ\>$ coeffcient. This was also the case in \cite{Dymarsky:2017xzb}.

\section{Results}
\label{sec:results}

\subsection{Universal bounds}
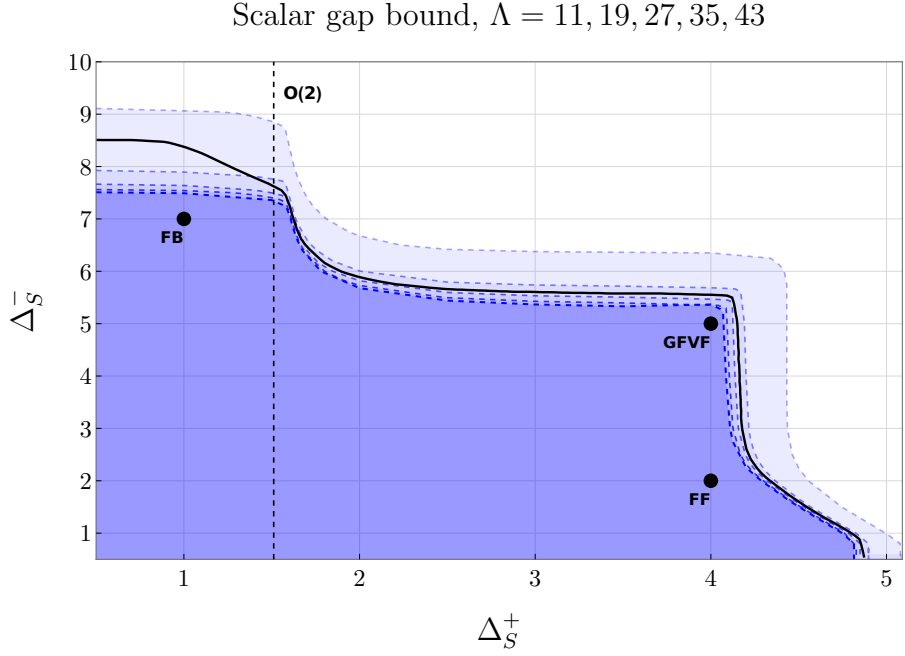
\begin{figure}
	\centering
	\begin{tikzpicture}
		\node[inner sep=0pt] (plot) {
			\includegraphics[width=0.8\textwidth]{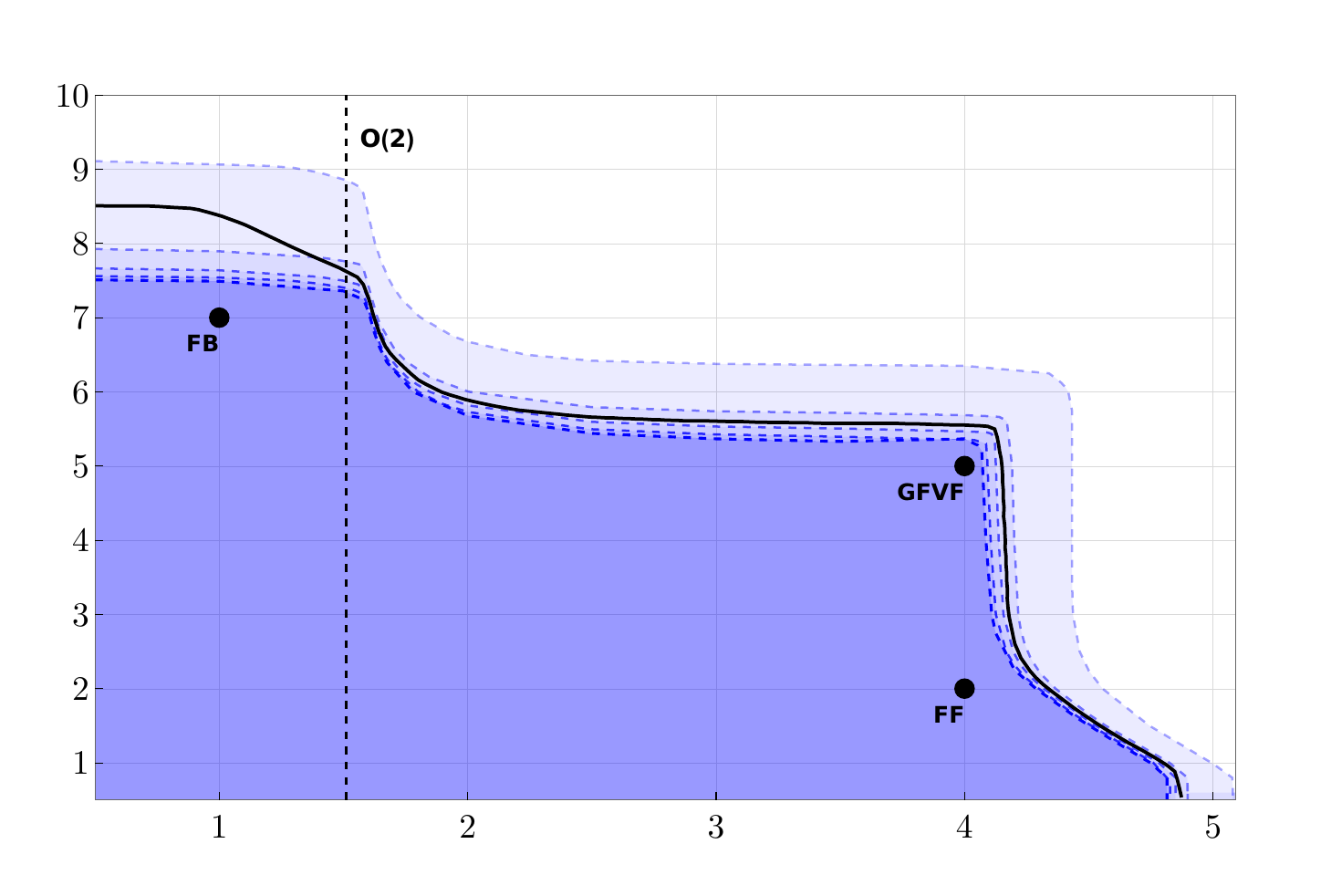}
		};

		\node[font=\large] at ($(plot.north)-(0,0.3)$)
		{Scalar gap bound, $\Lambda=11,19,27,35,43$};

		\node[font=\large] at ($(plot.south)+(0,-0.05)$)
		{$\Delta_S^+$};

		\node[font=\large, rotate=90] at ($(plot.west)+(-0.05,0)$)
		{$\Delta_S^-$};
	\end{tikzpicture}
	\caption{Bound on the allowed gaps for parity-even and parity-odd scalars in the $JT$ system. The bounds at derivative orders $\Lambda=11,19,27,35,43$ are shown. The solid black line shows the bound obtained using only the $\<JJJJ\>$ system at $\Lambda=23$ in \cite{Dymarsky:2017xzb}. The known values for gaps in the free complex boson (FB), generalized free vector field (GFVF), and free fermion (FF) are shown. The vertical dotted line is at the $\mathrm{O}(2)$ model value, $\Delta_S^+=1.51136$ from \cite{Chester:2019ifh}.}
	\label{fig:scalarGaps}
\end{figure}

As a first illustration of the mixed ($JT$) bootstrap setup, we present a universal bound on the scalar spectrum. As can be seen from the summary in appendix~\ref{ThreePointStructures}, the only scalar exchanges appearing in our setup are charge-conjugation-even scalars, with both spacetime parities allowed (note that there are no scalar operators present in $J\x T$ OPE).

We denote by $\Delta_S^+$ and $\Delta_S^-$ the dimensions of the lightest parity-even and parity-odd scalars, respectively, and study the resulting exclusion plot in the $(\Delta_S^+,\Delta_S^-)$ plane. From the discussion in section~\ref{sec:symmetry}, it follows that the resulting bound applies to the full system \eqref{eq:setnaive}, and not only to the symmetry-enhanced system \eqref{eq:setsmart}. It is shown in figure~\ref{fig:scalarGaps}.

This bound in theories with $\mathrm{U}(1)$ symmetry was first considered in the four-current bootstrap setup of \cite{Dymarsky:2017xzb}. Adding the stress tensor as an external operator gives a qualitatively similar bound in the region $\Delta_S^+\gtrsim 1.5$, while producing a stronger exclusion in the region of smaller $\Delta_S^+$ when compared at the same derivative cutoff $\Lambda$, see figure~\ref{fig:scalarGaps}. In this work we have also pushed the numerics further to $\Lambda=43$, where the bound appears to be reasonably well converged.
One might have hoped that the kink in the vicinity of the $\mathrm{O}(2)$ model sitting at $\Delta_S^+\approx 1.58$ would move towards the known $\mathrm{O}(2)$ value of $\Delta_S^+\approx 1.51136$ value with the higher derivative cutoff, but this does not appear to be the case.

\begin{figure}[t]
	\centering
	\begin{tikzpicture}
		\node[inner sep=0pt] (plot) {
			\includegraphics[width=1\textwidth]{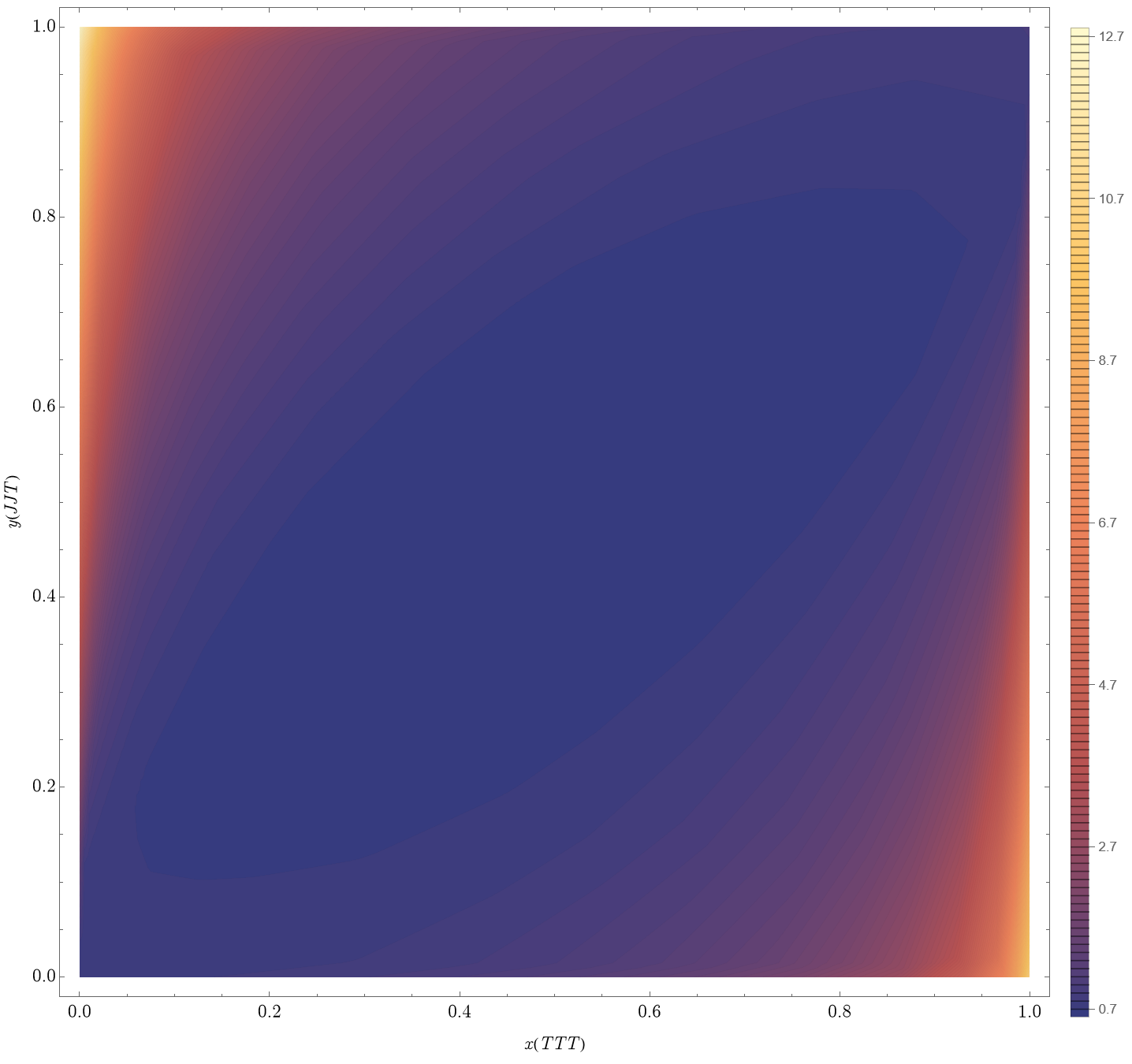}
		};

		\node[font=\large] at ($(plot.north)-(0,-0.2)$)
		{Lower bound on $\frac{c_T}{2c_B}$, $\Lambda=35$};


	\end{tikzpicture}
	\caption{Contour plot of lower bound on $\frac{c_T}{2c_B}$ at derivative order $\Lambda=35$, as a function of the parameters $x,y$ of the three-point functions $\<TTT\>$ and $\<JJT\>$.}
	\label{fig:centralCharge}
\end{figure}

In figure~\ref{fig:centralCharge} we present the universal lower bound on the central charge $c_T$ as a function of the $\<TTT\>$ and $\<JJT\>$ parameters $x$ and $y$; this bound applies to both systems~\eqref{eq:setnaive} and~\eqref{eq:setsmart}. Central bounds for the $\<TTTT\>$ system and the $\<JJJJ\>$ system were first individually considered in \cite{Dymarsky:2017yzx} (as a function of $x$) and \cite{Dymarsky:2017xzb} (as a function of $y$).
In the $x,y$ plane we can see a valley going diagonally around $x=y$ connecting the free boson ($x=y=0)$ and the free fermion ($x=y=1$). These lower bounds do not saturate any theories known to us. In particular, we expect the $\mathrm{O}(2)$ model to be close to be the free boson corner,
where the computed lower bound is well below the central charge of the $\mathrm{O}(2)$ model.
The lower bound increases as we move away from this valley, with the lower bound blowing up at $(x,y)=(0,1)$ and $(1,0)$.

\begin{figure}
	\centering
	\begin{tikzpicture}
		\node[inner sep=0pt] (plot) {
			\includegraphics[width=1\textwidth]{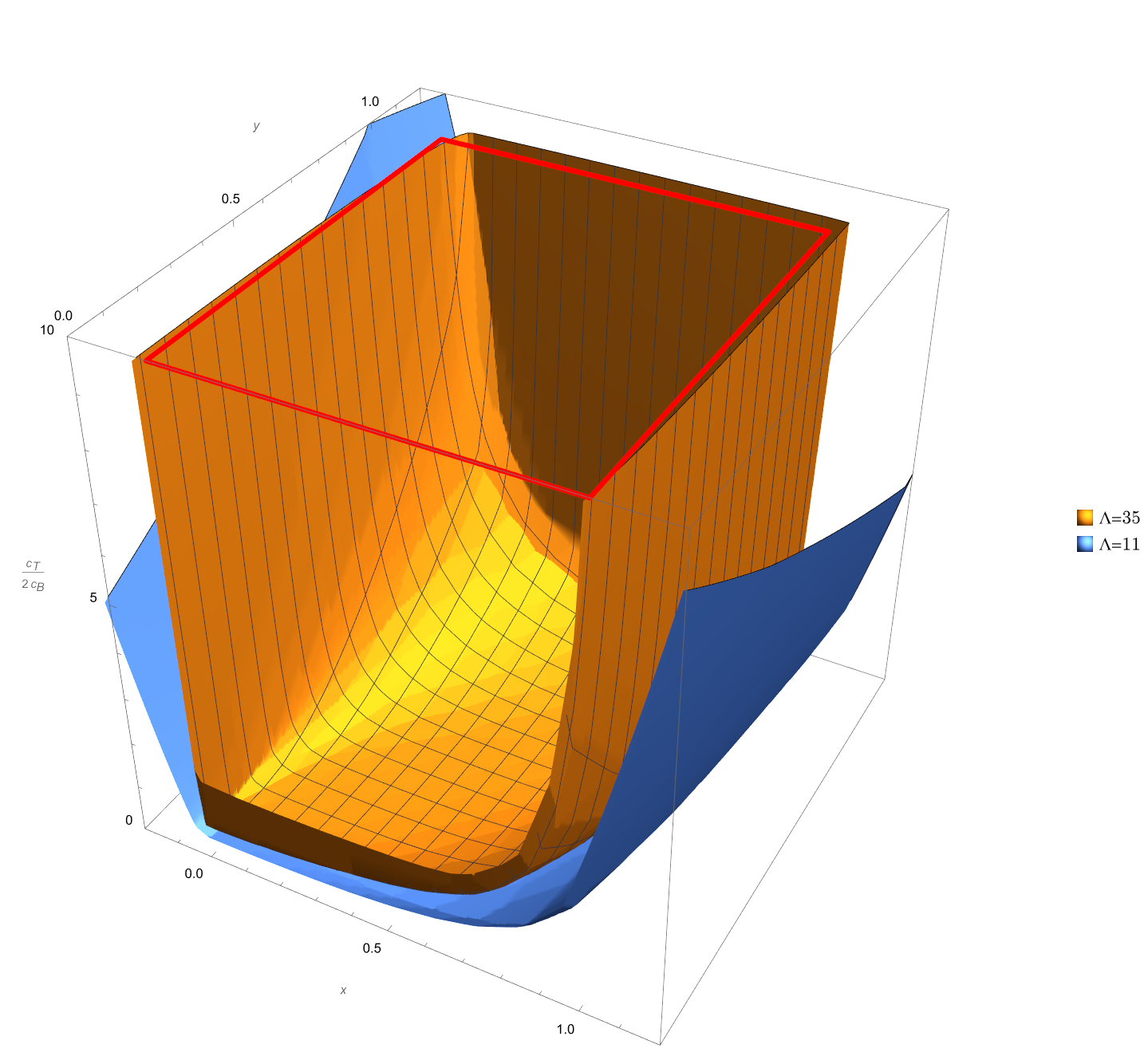}
		};

		\node[font=\large] at ($(plot.north)-(0,-0.1)$)
		{Lower bound on $\frac{c_T}{2c_B}$, $\Lambda=11,35$};


	\end{tikzpicture}
	\caption{Lower bound on $\frac{c_T}{2c_B}$, including points slightly outside the conformal collider bounds. The red square shows the region defined by the conformal collider bounds~\eqref{eq:HMbounds}. For clarity, the plot is deliberately cut off at the front face. Increasing $\L$ from 11 to 35 shows rapid growth of the lower bound outside of the region defined by the conformal collider bounds.}
	\label{fig:conformalColliderBound}
\end{figure}

If figure~\ref{fig:conformalColliderBound} we show that, similarly to~\cite{Dymarsky:2017yzx}, the $c_T$ lower bound reproduces the conformal collider bounds~\eqref{eq:HMbounds} as $\L$ is increased. It appears that our bounds may be slightly stronger near the corners where $(x,y)$ is $(0,1)$ and $(1,0)$, which can also be seen in figure~\ref{fig:centralCharge}.

\begin{figure}
	\centering

	\begin{minipage}[t]{0.48\textwidth}
		\centering
		\begin{tikzpicture}
			\node[inner sep=0pt] (plotA) {
				\includegraphics[height=.9\textwidth]{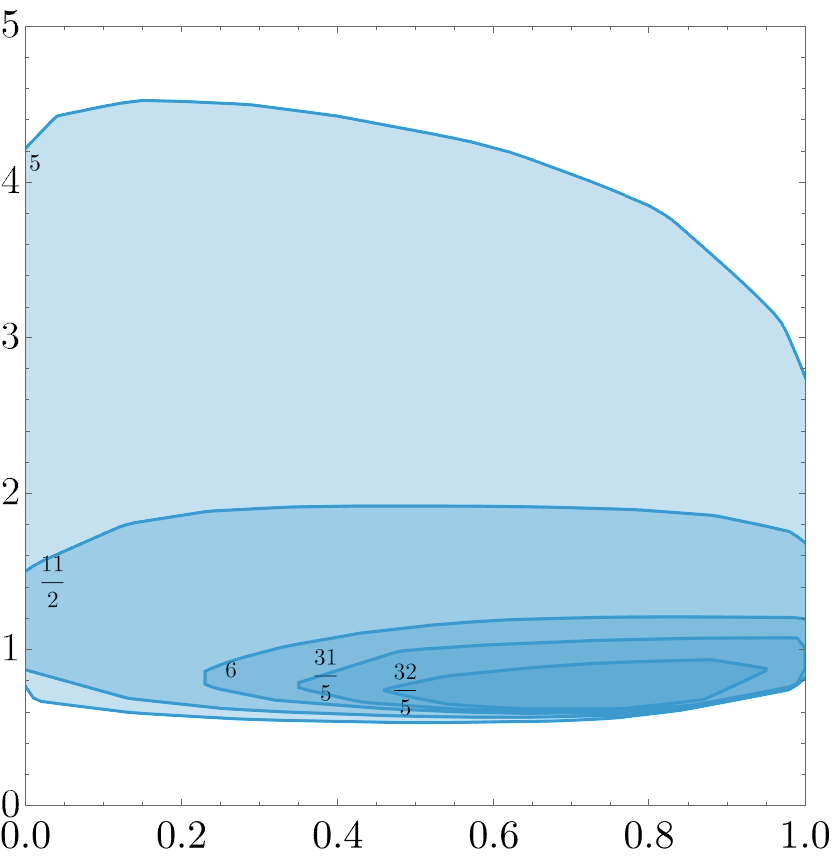}
			};

			\node[font=\small] at ($(plotA.north)+(4.2,0.25)$)
			{$c_T$ bounds, $\Delta_2 \geq 5.0,\dots 6.4$};

			\node[font=\small] at ($(plotA.south)+(0,-0.25)$)
			{$x(TTT)$};

			\node[font=\small, rotate=90] at ($(plotA.west)+(-0.5,0)$)
			{$\frac{c_T}{2c_B}$};
		\end{tikzpicture}
	\end{minipage}
	\hfill
	\begin{minipage}[t]{0.48\textwidth}
		\centering
		\begin{tikzpicture}
			\node[inner sep=0pt] (plotB) {
				\includegraphics[height=.9\textwidth]{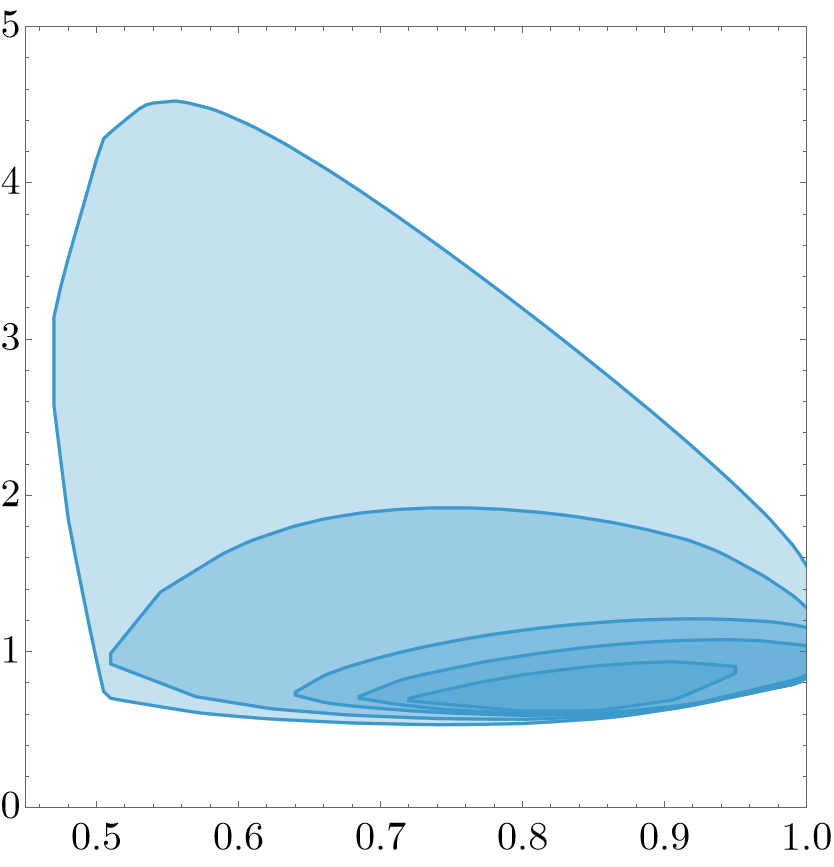}
			};

			\node[font=\small] at ($(plotB.north)+(0,0.25)$)
			{};

			\node[font=\small] at ($(plotB.south)+(0,-0.25)$)
			{$y(JJT)$};

			\node[font=\small, rotate=90] at ($(plotB.west)+(-0.5,0)$)
			{$\frac{c_T}{2c_B}$};
		\end{tikzpicture}
	\end{minipage}

	\caption{Closed allowed regions in the $(x,y,c_T)$ volume for various values of gaps in the spin-2 parity-even channel. The shadow of the volumes on the $(x,c_T)$ (left) and the $(y,c_T)$ (right) planes are shown.}
	\label{fig:cTspin2Gaps}
\end{figure}

There is no upper bound on $c_T$ without additional assumptions. This is due to theories with infinite central charge, like generalized free fields (of $T$ and $J$) and large $N$ theories. Such theories can be excluded by including gap assumptions.\footnote{Note that usually the absence of an upper $c_T$ bound is a special case of the general principle that numerical bootstrap cannot put lower bounds on OPE coefficients ($1/c_T$ enters as an OPE coefficient squared) without isolating the operator of interest by gap assumptions. This is not the case here: $\<TTT\>$ has more structures than a generic spin-2 exchange, and that makes $T$ effectively isolated. The gaps that we impose here play the role of excluding large-$c_T$ theories. This can be achieved in various other ways, e.g.\ by scalar gaps~\cite{Dymarsky:2017yzx}.} Here, we present central charge bounds with a gap assumption in the spin-2 parity even spectrum, shown in figure~\ref{fig:cTspin2Gaps}. Note that we impose the same gap in both the charge-conjugation-even and charge-conjugation-odd sectors, and so this bound applies to~\eqref{eq:setnaive} as well as~\eqref{eq:setsmart} systems.
Imposing a high enough gap restricts the allowed values of $x$ and $y$ to an island that shrinks as we increase the gap. Additionally, the upper bound on the central charge decreases quickly with the gap, while the lower bound increases very little to around 0.62.

\subsection{\texorpdfstring{$\mathrm{O}(2)$}{\mathrm{O}(2)} model}
\label{sec:O2}

\begin{figure}
	\centering

	\begin{minipage}[t]{0.48\textwidth}
		\centering
		\begin{tikzpicture}
			\node[inner sep=0pt] (plotA) {
				\includegraphics[height=.9\textwidth]{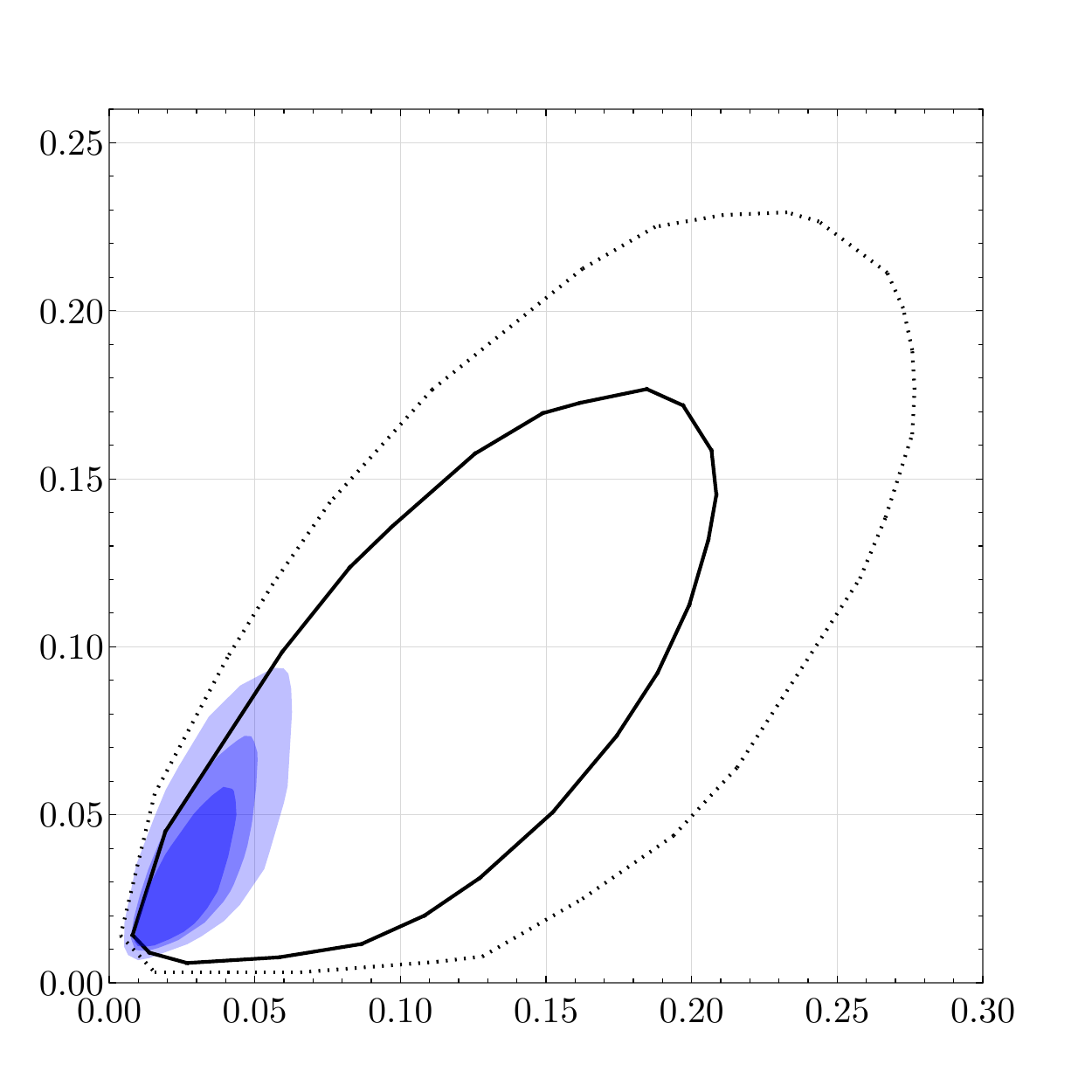}
			};

			\node[font=\small] at ($(plotA.north)+(0,0.25)$)
			{$\mathrm{O}(2)$ model, $\Lambda=19,27,35$};

			\node[font=\small] at ($(plotA.south)+(0,-0.05)$)
			{$x(TTT)$};

			\node[font=\small, rotate=90] at ($(plotA.west)+(-0.05,0)$)
			{$y(JJT)$};
		\end{tikzpicture}
	\end{minipage}
	\hfill
	\begin{minipage}[t]{0.48\textwidth}
		\centering
		\begin{tikzpicture}
			\node[inner sep=0pt] (plotB) {
				\includegraphics[height=.9\textwidth]{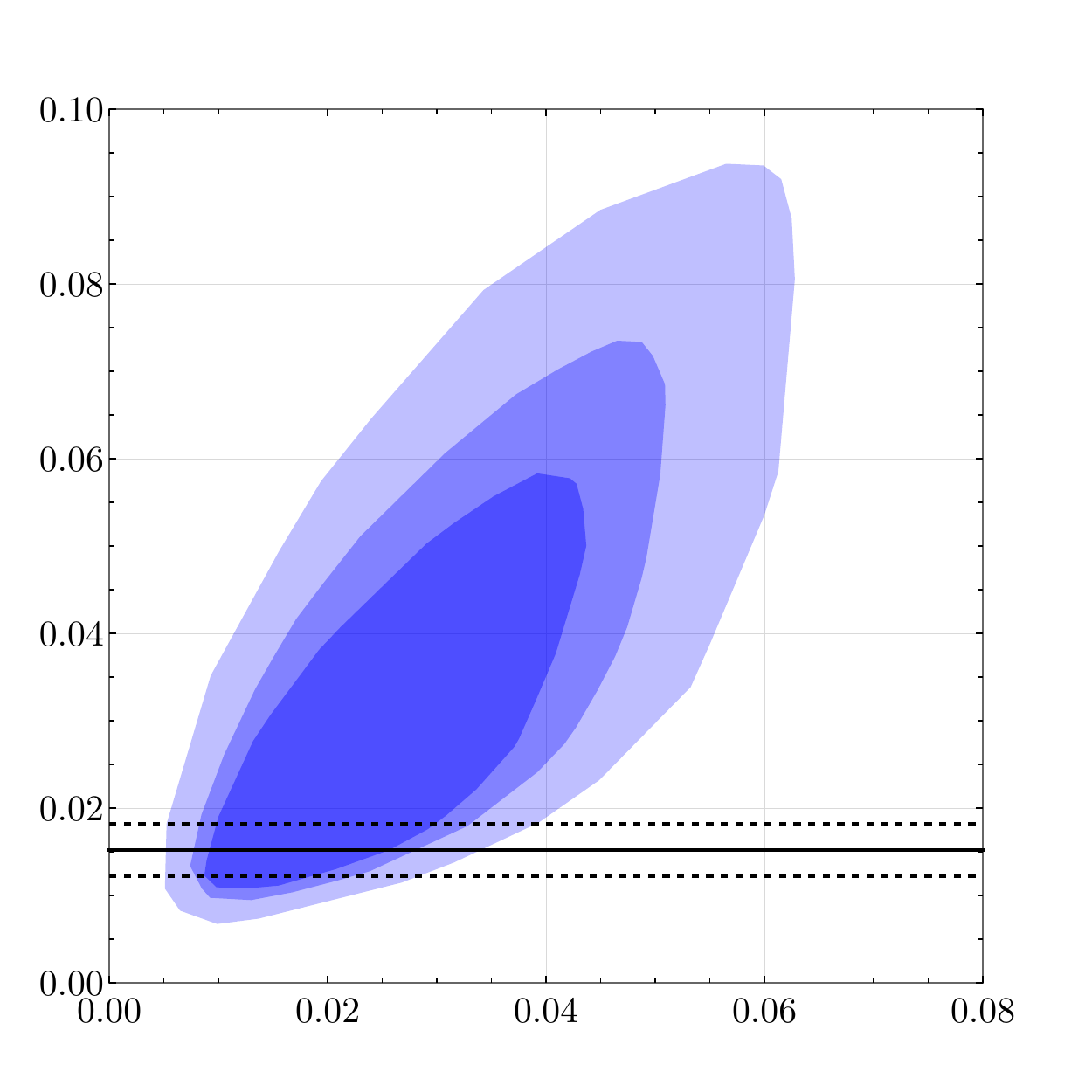}
			};

			\node[font=\small] at ($(plotB.north)+(0,0.25)$)
			{Zoom into stronger-assumption island};

			\node[font=\small] at ($(plotB.south)+(0,-0.05)$)
			{$x(TTT)$};

			\node[font=\small, rotate=90] at ($(plotB.west)+(-0.05,0)$)
			{$y(JJT)$};
		\end{tikzpicture}
	\end{minipage}

	\caption{Allowed regions in the $(x,y)$ plane, using the $\mathrm{O}(2)$ model assumptions in Table~\ref{tbl:O2}. Left: The dotted and solid black lines show the bound given by the weaker assumptions at $\Lambda=19,27$. The blue region represents the bound given by the stronger assumptions at $\Lambda=19,27,35$. Right: zoom into the stronger-assumption island. Horizontal lines denotes the value and uncertainty of the $\mathrm{O}(2)$ prediction for $\gamma_{JJT}$ from Ref.~\cite{Reehorst:2019pzi}.}
	\label{fig:o2Model}
\end{figure}

We now consider the $\mathrm{O}(2)$ model as a natural target for the $JT$ setup. One useful feature of this example is that the $\mathrm{O}(2)$ model is invariant under the $\mathbb Z_2$ charge conjugation symmetry $C$ described in section~\ref{sec:symmetry}: it is a reflection in the $\mathrm{O}(2)$ group. In other words, $C$-parity is equivalent to $\mathrm{O}(2)$ parity. In our setup, we have access to $\mathrm{O}(2)$-spin 0 operators of both $C$-parities. We will only interpret the bounds presented in this section in the context of the $\mathrm{O}(2)$ model and the $C$-symmetric system~\eqref{eq:setsmart}.

Our targets are the parameters $\gamma_{JJT}$ and $\gamma_{TTT}$ which are related to the OPE coefficients $\lambda_{JJT}$ and $\lambda_{TTT}$ as described in section~\ref{sec:WardIds}. We present out plots using the $x,y$ variables defined in \eqref{eq:xy}. In figure~\ref{fig:o2Model} we plot the islands for the $\mathrm{O}(2)$ model in $x,y$ plane, obtained by imposing two sets of assumptions that we refer to as the weaker and the stronger sets.

Both sets of assumptions fix the values of the scaling dimension $\De_S^+$ of the lightest parity-even $C$-even scalar and the value of $c_T$ to those obtained for the $\mathrm{O}(2)$ model in~\cite{Chester:2019ifh}.\footnote{Note that the lightest parity-even $C$-even scalar in the $\mathrm{O}(2)$ model is $s=\phi^2$.} Furthermore, in both cases we make an assumption about $\De_{S'}^+$, the dimension of the second-lightest parity-even $C$-even scalar. The stronger set of assumptions additionally constrains the lightest parity-odd $C$-even scalar and the first parity-even $C$-even spin-2 operator above $T$. All our assumptions are summarized in table~\ref{tbl:O2}. We now provide some justification for these choices.

We consider the weaker set of assumptions, those only on $c_T$, $\De_S^+$, $\De_{S'}^+$, to be rather conservative. As mentioned above, the values of $c_T$, $\De_S^+$ are based on the bootstrap determinations in~\cite{Chester:2019ifh}. The assumption ${\Delta_S^+}'>3.75$ is based on the bootstrap (${\Delta_S^+}'=3.794(8)$ in~\cite{Liu:2020tpf}), Monte-Carlo (${\Delta_S^+}'=3.789(4)$ in~\cite{Hasenbusch:2019wla}), and Fuzzy-Sphere (${\Delta_S^+}'=3.887999$\footnote{This is direct exact diagonalization result with $N=13$ electrons, provided without an explicit error estimate in~\cite{Dey:2026cso}.\label{footnote:EDresult}} in~\cite{Dey:2026cso}). The island obtained with these assumptions is shown in the left panel of figure~\ref{fig:o2Model}.

The stronger set of assumptions contains $\Delta_S^->6$.
This is motivated by the expectation that the lightest parity-odd scalar in the $\mathrm{O}(2)$ model is relatively heavy. In the free complex scalar theory the corresponding operator has dimension $\Delta_S^-=7$. We expect the interacting $\mathrm{O}(2)$ model to be not far from the kink in figure~\ref{fig:scalarGaps} that also sits above~$\De_S^-=7$. Finally, no evidence of a parity-odd primary state is seen below $\De=6$ in the Fuzzy-Sphere data presented in~\cite{Dey:2026cso}.\footnote{Note that space parity is not a conserved quantum number in their microscopic realization.}

Furthermore, the stronger set of assumptions includes $\Delta_{T'}>4$.
Here $T'$ denotes the first parity-even $C$-even spin-2 primary above the conserved stress tensor. This is based on the Fuzzy-Sphere estimate $\De_{T'}=4.596819^{\ref{footnote:EDresult}}$ in~\cite{Dey:2026cso}. The value $\De_{T'}$ is also supported by $4-\varepsilon$-expansion calculations in~\cite{Henriksson:2025vyi}.

Using the stronger assumptions we obtain the island in figure~\ref{fig:o2Model} at $\Lambda=19,27,35$ and determine
\be
x = 0.026 (18), \quad \gamma_{TTT}=-0.0790(29),\nn\\
y = 0.035 (24),\quad\gamma_{JJT}=-0.0776(40).
\ee
If we assume the $\gamma_{JJT}$ result of \cite{Reehorst:2019pzi}, we obtain
\be
x= 0.019 (11),\quad \gamma_{TTT}=-0.0802(17).
\ee

\begin{table}
	\centering
	\renewcommand{\arraystretch}{1.18}
	\begin{tabular}{c c c}
		\hline
		\multicolumn{3}{c}{$\mathrm{O}(2)$ assumptions} \\
		\hline
		                        & Weaker    & Stronger  \\
		\hline
		$\Delta_S^+$            & $1.51136$ & $1.51136$ \\
		$c_T/c_T^{\text{free}}$ & $0.944$   & $0.944$   \\
		\hline
		$\Delta_{S'}^+$         & $>3.75$   & $>3.75$   \\
		$\Delta_S^-$            & --        & $>6$      \\
		$\Delta_{T'}$           & --        & $>4$      \\
		\hline
	\end{tabular}
	\caption{Assumptions used to isolate the $\mathrm{O}(2)$ model. The weaker setup fixes the leading parity-even scalar dimension and the stress-tensor central charge, and imposes a gap to the next parity-even scalar $S'$. The stronger setup additionally imposes a gap to the lightest parity-odd scalar $S^-$ and a gap to the next spin-$2$, parity-even, $\mathbb{Z}_2$-even operator $T'$ above the stress tensor. The superscript $\pm$ denotes spacetime parity.}
	\label{tbl:O2}
\end{table}

\section{Discussion}
\label{sec:discussion}

This work follows the spirit of \cite{Dymarsky:2017yzx,Dymarsky:2017xzb} in exploring universal bootstrap bounds on the space of local CFTs, as opposed to focusing on a single or small set of theories. Some of our key results are an improved bound on parity even/odd scalar operator gaps in any 3d CFT with a $U(1)$ symmetry, and general bounds on stress tensor and current OPE coefficients. There are several natural extensions that are worth exploring.

Firstly, the bootstrap system studied in this paper provides a useful warmup for studying correlation functions in the $O(2)$ model that include currents and stress tensors. The success of previous work studying stress tensor correlators in the 3d Ising CFT \cite{Chang:2024whx} indicates that this could be a very powerful way to obtain precise results about the $O(2)$ model. However, there are some choices to be made in this approach. Previous bootstrap studies of the $O(2)$ model have focused on successively larger systems of scalar correlators --- first correlators of the charge-1 scalar $\phi$ \cite{Kos:2013tga}, then mixed correlators including $\phi$ and the singlet $s$ \cite{Kos:2015mba,Kos:2016ysd}, then correlators including $\phi,s$, and the leading charge-2 scalar $t$ \cite{Chester:2019ifh}. Each larger system of correlators led to stronger bounds, for example eventually yielding a clear judgement on the tension between Monte Carlo simulations and experimental results in the $O(2)$ model in \cite{Chester:2019ifh}.

In an ideal world, we would simply add $T_{\mu\nu}$ and $J_\mu$ to the mix, studying a system of correlation functions of $\phi,s,t,T_{\mu\nu},J_{\mu}$ in order to obtain strictly stronger bounds. Unfortunately, this full system of correlation functions may be too large for current computational resources. There are two glaring issues with larger systems: firstly, the larger number of correlation functions (including multiple four-point structures) means that the bootstrap functionals have more components, making the resulting SDPs more expensive. Secondly, more OPE coefficients between ``external" operators (i.e.~operators whose four-point functions we are studying) appear in the crossing equations, meaning one must scan over a higher dimensional space of OPE coefficients to compute bounds. This scan can potentially be handled with the ``cutting surface" algorithm of \cite{Chester:2019ifh}, or navigator function methods \cite{Reehorst:2021ykw,Reehorst:2021hmp,Su:2022xnj,Sirois:2022vth}, but higher dimensionality necessarily makes the scan more expensive and difficult.

As a preliminary step, we can restrict to a subset of the operators $\phi,s,t,T_{\mu\nu},J_{\mu}$, and obtain novel, but not necessarily strictly stronger, bounds. We expect it should be computationally straightforward to study correlators of $\phi, T_{\mu\nu},J_{\mu}$ by slight modifications of the methods in this work. It might also be in reach to include $s$ with present resources. However, we do not have a good understanding of how these tradeoffs will affect the strength of the resulting bounds. The ``large number of correlators'' frontier is mostly unexplored in the numerical conformal bootstrap, and these investigations would play the role of initial steps.

Another natural extension of this work is to generalize $U(1)$ to other Lie groups --- in particular nonabelian groups, extending the study of nonabelian currents in \cite{He:2023ewx} to include stress tensors. A virtue of crossing equations for nonabelian groups is that $c_J$ (the current two-point function) appears nontrivially, so we should expect to obtain universal bounds on $c_J$. (By contrast, in the abelian case, $c_J$ dropped out of the bootstrap equations that we studied.) The weak gravity conjecture \cite{Arkani-Hamed:2006emk} implies that, in holographic theories, there should be an upper bound on the ratio $c_J/c_T$, and it would be interesting to explore this possibility from the numerical bootstrap. Note that bounds relating $c_J$ and $c_T$ have already been explored in some supersymmetric contexts, where the stress tensor and flavor currents can be embedded in SUSY multiplets with scalar super-primaries, see e.g.\ \cite{Beem:2014zpa}.

As far as we are aware, little is known about the space of non-supersymmetric CFTs with nonabelian symmetries, particularly for exotic Lie groups. For example, one can ask: what is the simplest 3d CFT with $E_8$ symmetry? Bootstrap bounds could provide important clues about such a theory. At the same time, it would be interesting to explore this question with non-bootstrap methods, for example by searching for critical points with $E_8$ symmetry using lattice or fuzzy sphere techniques \cite{Zhu:2022gjc}.\footnote{We thank John McGreevy for discussion of this idea.}

In this work, we focused on CFTs with parity symmetry; another interesting extension would be to consider theories with parity violation, including, for example, Chern-Simons-matter theories. It would also be interesting to see whether including correlators of currents and stress tensors can help isolate theories that have been more elusive to bootstrap techniques, like QED$_3$ \cite{Li:2018lyb,Li:2020bnb,Albayrak:2021xtd,Li:2021emd}.

\section*{Acknowledgements}

We thank Christopher Herzog, John McGreevy, Matthew Mitchell, Hirosi Ooguri, Mayur Sharma, Cumrun Vafa, and Naveen Umasankar for discussions. We thank Yixin Xu for many discussions and collaboration in the initial stages of this work.
The work of RSE is supported by Simons Foundation grant 915279 (IHES), Department of Energy grant DE-SC0007859, and a Leinweber Postdoctoral Fellowship from the University of Michigan. The work of PK funded by UK Research and Innovation (UKRI) under the UK
government’s Horizon Europe funding Guarantee [grant number EP/X042618/1] and by Science and Technology Facilities Council [grant number ST/X000753/1].
This material is based upon work supported by the U.S.\ Department of Energy, Office of Science, Office of
High Energy Physics, under Award Numbers DE-SC0011632 and DE-SC0017660. This work used the Expanse cluster at the San Diego Supercomputing Center (SDSC) through allocation PHY190023 from the Advanced Cyberinfrastructure Coordination Ecosystem: Services \& Support (ACCESS) program, which is supported by National Science Foundation grants \#2138259, \#2138286, \#2138307, \#2137603, and \#2138296.

\newpage

\appendix
\section{Crossing equations}

\label{ap:crossing}
In this appendix we use the cross-ratios  $x = ({z +\bar{z}})/{2}, t = ({z-\bar{z}})/{2i}$.
The crossing equations are written in terms of the derivatives
$\left.\partial_x^m\partial_t^n \left(g_{\mathcal{O}_1\mathcal{O}_2\mathcal{O}_3\mathcal{O}_4}^{{\langle q_1 q_2 q_3 q_4 \rangle}^\pm}\right)\right\vert_*$,
where $\vert_*$ denotes evaluation at the crossing symmetric point $x=1$, $t=0$.
With our choice of tensor structures, only even $n$ are needed, while the parity of $m$ is fixed by
the signature of the corresponding structure under the $(13)$ permutation.
In the numerical study, the number of equations used is determined by $\Lambda$ through the constraint $m + 2n \leq \Lambda$.

\vspace{.25cm}

{\allowdisplaybreaks[1]
	\noindent\textbf{Bulk equations for $\langle JJJJ\rangle$}
\begin{alignat*}{2}
0 &=
\partial_x^m \partial_t^n
\left.
\left(
g_{JJJJ}^{\langle -1,-1,-1,-1\rangle^+}
\right)
\right|_{*},
&\qquad& (m\ \mathrm{odd})
\\
0 &=
\partial_x^m \partial_t^n
\left.
\left(
g_{JJJJ}^{\langle 1,-1,-1,-1\rangle^+}
\right)
\right|_{*},
&\qquad& (m\ \mathrm{odd})
\\
0 &=
\partial_x^m \partial_t^n
\left.
\left(
g_{JJJJ}^{\langle 0,0,-1,-1\rangle^+}
    + g_{JJJJ}^{\langle -1,0,0,-1\rangle^+}
\right)
\right|_{*},
&\qquad& (m\ \mathrm{odd})
\\
0 &=
\partial_x^m \partial_t^n
\left.
\left(
g_{JJJJ}^{\langle 0,0,-1,-1\rangle^+}
    - g_{JJJJ}^{\langle -1,0,0,-1\rangle^+}
\right)
\right|_{*},
&\qquad& (m\ \mathrm{even})
\\
0 &=
\partial_x^m \partial_t^n
\left.
\left(
g_{JJJJ}^{\langle 1,1,-1,-1\rangle^+}
    + g_{JJJJ}^{\langle -1,1,1,-1\rangle^+}
\right)
\right|_{*},
&\qquad& (m\ \mathrm{odd})
\\
\intertext{\noindent\textbf{Line equations for $\langle JJJJ\rangle$}}
0 &=
\partial_x^m
\left.
\left(
g_{JJJJ}^{\langle 1,1,-1,-1\rangle^+}
    - g_{JJJJ}^{\langle -1,1,1,-1\rangle^+}
\right)
\right|_{*},
&\qquad& (m\ \mathrm{even})
\\
0 &=
\partial_x^m
\left.
\left(
g_{JJJJ}^{\langle 1,0,0,-1\rangle^+}
    + g_{JJJJ}^{\langle 0,0,1,-1\rangle^+}
\right)
\right|_{*},
&\qquad& (m\ \mathrm{odd})
\end{alignat*}

	\noindent\textbf{Bulk equations for $\langle TTTT\rangle$}
\begin{alignat*}{2}
0 &=
\partial_x^m \partial_t^n
\left.
\left(
g_{TTTT}^{\langle -2,-2,-2,-2\rangle^+}
\right)
\right|_{*},
&\qquad& (m\ \mathrm{odd})
\\
0 &=
\partial_x^m \partial_t^n
\left.
\left(
g_{TTTT}^{\langle 0,-2,-2,-2\rangle^+}
\right)
\right|_{*},
&\qquad& (m\ \mathrm{odd})
\\
0 &=
\partial_x^m \partial_t^n
\left.
\left(
g_{TTTT}^{\langle 2,-2,-2,-2\rangle^+}
\right)
\right|_{*},
&\qquad& (m\ \mathrm{odd})
\\
0 &=
\partial_x^m \partial_t^n
\left.
\left(
g_{TTTT}^{\langle -1,-1,-2,-2\rangle^+}
    + g_{TTTT}^{\langle -2,-1,-1,-2\rangle^+}
\right)
\right|_{*},
&\qquad& (m\ \mathrm{odd})
\\
0 &=
\partial_x^m \partial_t^n
\left.
\left(
g_{TTTT}^{\langle -1,-1,-2,-2\rangle^+}
    - g_{TTTT}^{\langle -2,-1,-1,-2\rangle^+}
\right)
\right|_{*},
&\qquad& (m\ \mathrm{even})
\\
\intertext{\noindent\textbf{Line equations for $\langle TTTT\rangle$}}
0 &=
\partial_x^m
\left.
\left(
g_{TTTT}^{\langle 1,-1,-2,-2\rangle^+}
    + g_{TTTT}^{\langle -2,-1,1,-2\rangle^+}
\right)
\right|_{*},
&\qquad& (m\ \mathrm{odd})
\\
0 &=
\partial_x^m
\left.
\left(
g_{TTTT}^{\langle 1,-1,-2,-2\rangle^+}
    - g_{TTTT}^{\langle -2,-1,1,-2\rangle^+}
\right)
\right|_{*},
&\qquad& (m\ \mathrm{even})
\\
0 &=
\partial_x^m
\left.
\left(
g_{TTTT}^{\langle 0,0,-2,-2\rangle^+}
    + g_{TTTT}^{\langle -2,0,0,-2\rangle^+}
\right)
\right|_{*},
&\qquad& (m\ \mathrm{odd})
\\
0 &=
\partial_x^m
\left.
\left(
g_{TTTT}^{\langle 2,0,-2,-2\rangle^+}
    + g_{TTTT}^{\langle -2,0,2,-2\rangle^+}
\right)
\right|_{*},
&\qquad& (m\ \mathrm{odd})
\\
0 &=
\partial_x^m
\left.
\left(
g_{TTTT}^{\langle 2,0,-2,-2\rangle^+}
    - g_{TTTT}^{\langle -2,0,2,-2\rangle^+}
\right)
\right|_{*},
&\qquad& (m\ \mathrm{even})
\\
0 &=
\partial_x^m
\left.
\left(
g_{TTTT}^{\langle 1,1,-2,-2\rangle^+}
    + g_{TTTT}^{\langle -2,1,1,-2\rangle^+}
\right)
\right|_{*},
&\qquad& (m\ \mathrm{odd})
\\
0 &=
\partial_x^m
\left.
\left(
g_{TTTT}^{\langle 2,2,-2,-2\rangle^+}
    + g_{TTTT}^{\langle -2,2,2,-2\rangle^+}
\right)
\right|_{*},
&\qquad& (m\ \mathrm{odd})
\\
0 &=
\partial_x^m
\left.
\left(
g_{TTTT}^{\langle 2,2,-2,-2\rangle^+}
    - g_{TTTT}^{\langle -2,2,2,-2\rangle^+}
\right)
\right|_{*},
&\qquad& (m\ \mathrm{even})
\\
0 &=
\partial_x^m
\left.
\left(
g_{TTTT}^{\langle 2,1,-1,-2\rangle^+}
    + g_{TTTT}^{\langle -1,1,2,-2\rangle^+}
\right)
\right|_{*},
&\qquad& (m\ \mathrm{odd})
\\
\intertext{\noindent\textbf{Point equations for $\langle TTTT\rangle$}}
0 &=
\left.
\left(
g_{TTTT}^{\langle 0,0,-2,-2\rangle^+}
    - g_{TTTT}^{\langle -2,0,0,-2\rangle^+}
\right)
\right|_{*},
&\qquad&
\\
0 &=
\left.
\left(
g_{TTTT}^{\langle 1,1,-2,-2\rangle^+}
    - g_{TTTT}^{\langle -2,1,1,-2\rangle^+}
\right)
\right|_{*},
&\qquad&
\\
0 &=
\left.
\left(
g_{TTTT}^{\langle 1,0,-1,-2\rangle^+}
    - g_{TTTT}^{\langle -1,0,1,-2\rangle^+}
\right)
\right|_{*},
&\qquad&
\\
0 &=
\left.
\left(
g_{TTTT}^{\langle 2,1,-1,-2\rangle^+}
    - g_{TTTT}^{\langle -1,1,2,-2\rangle^+}
\right)
\right|_{*}.
&\qquad&
\end{alignat*}

	\noindent\textbf{Bulk equations for $\langle TJTJ\rangle$}
\begin{alignat*}{2}
0 &=
\partial_x^m \partial_t^n
\left.
\left(
g_{TJTJ}^{\langle -2,-1,-2,-1\rangle^+}
\right)
\right|_{*},
&\qquad& (m\ \mathrm{odd})
\\
0 &=
\partial_x^m \partial_t^n
\left.
\left(
g_{TJTJ}^{\langle -2,1,-2,-1\rangle^+}
\right)
\right|_{*},
&\qquad& (m\ \mathrm{odd})
\\
0 &=
\partial_x^m \partial_t^n
\left.
\left(
g_{TJTJ}^{\langle -2,0,-2,0\rangle^+}
\right)
\right|_{*},
&\qquad& (m\ \mathrm{odd})
\\
0 &=
\partial_x^m \partial_t^n
\left.
\left(
g_{TJTJ}^{\langle -2,1,-2,1\rangle^+}
\right)
\right|_{*},
&\qquad& (m\ \mathrm{odd})
\\
0 &=
\partial_x^m \partial_t^n
\left.
\left(
g_{TJTJ}^{\langle -1,0,-2,-1\rangle^+}
    - g_{TJTJ}^{\langle -2,0,-1,-1\rangle^+}
\right)
\right|_{*},
&\qquad& (m\ \mathrm{even})
\\
0 &=
\partial_x^m \partial_t^n
\left.
\left(
g_{TJTJ}^{\langle 0,-1,-2,-1\rangle^+}
\right)
\right|_{*},
&\qquad& (m\ \mathrm{odd})
\\
\intertext{\noindent\textbf{Line equations for $\langle TJTJ\rangle$}}
0 &=
\partial_x^m
\left.
\left(
g_{TJTJ}^{\langle -1,1,-2,0\rangle^+}
    - g_{TJTJ}^{\langle -2,1,-1,0\rangle^+}
\right)
\right|_{*},
&\qquad& (m\ \mathrm{even})
\\
0 &=
\partial_x^m
\left.
\left(
g_{TJTJ}^{\langle 0,1,-2,-1\rangle^+}
    + g_{TJTJ}^{\langle -2,1,0,-1\rangle^+}
\right)
\right|_{*},
&\qquad& (m\ \mathrm{odd})
\\
0 &=
\partial_x^m
\left.
\left(
g_{TJTJ}^{\langle 0,0,-2,0\rangle^+}
\right)
\right|_{*},
&\qquad& (m\ \mathrm{odd})
\\
0 &=
\partial_x^m
\left.
\left(
g_{TJTJ}^{\langle 0,1,-2,1\rangle^+}
\right)
\right|_{*},
&\qquad& (m\ \mathrm{odd})
\\
0 &=
\partial_x^m
\left.
\left(
g_{TJTJ}^{\langle 1,1,-2,0\rangle^+}
    - g_{TJTJ}^{\langle -2,1,1,0\rangle^+}
\right)
\right|_{*},
&\qquad& (m\ \mathrm{even})
\\
\intertext{\noindent\textbf{Point equations for $\langle TJTJ\rangle$}}
0 &=
\left.
\left(
g_{TJTJ}^{\langle 1,0,-2,-1\rangle^+}
    - g_{TJTJ}^{\langle -2,0,1,-1\rangle^+}
\right)
\right|_{*}.
&\qquad&
\end{alignat*}

	\noindent\textbf{Bulk equations for $\langle JTTJ\rangle$ and $\langle TTJJ\rangle$}
\begin{alignat*}{2}
0 &=
\partial_x^m \partial_t^n
\left.
\left(
g_{JTTJ}^{\langle -1,-2,-2,-1\rangle^+}
    + g_{TTJJ}^{\langle -2,-2,-1,-1\rangle^+}
\right)
\right|_{*},
&\qquad& (m\ \mathrm{even})
\\
0 &=
\partial_x^m \partial_t^n
\left.
\left(
g_{JTTJ}^{\langle -1,-2,-2,-1\rangle^+}
    + g_{TTJJ}^{\langle -2,-2,-1,-1\rangle^+}
\right)
\right|_{*},
&\qquad& (m\ \mathrm{odd})
\\
0 &=
\partial_x^m \partial_t^n
\left.
\left(
g_{JTTJ}^{\langle 1,-2,-2,-1\rangle^+}
    + g_{TTJJ}^{\langle -2,-2,1,-1\rangle^+}
\right)
\right|_{*},
&\qquad& (m\ \mathrm{even})
\\
0 &=
\partial_x^m \partial_t^n
\left.
\left(
g_{JTTJ}^{\langle 1,-2,-2,-1\rangle^+}
    + g_{TTJJ}^{\langle -2,-2,1,-1\rangle^+}
\right)
\right|_{*},
&\qquad& (m\ \mathrm{odd})
\\
0 &=
\partial_x^m \partial_t^n
\left.
\left(
g_{JTTJ}^{\langle 0,-2,-2,0\rangle^+}
    + g_{TTJJ}^{\langle -2,-2,0,0\rangle^+}
\right)
\right|_{*},
&\qquad& (m\ \mathrm{even})
\\
0 &=
\partial_x^m \partial_t^n
\left.
\left(
g_{JTTJ}^{\langle 0,-2,-2,0\rangle^+}
    + g_{TTJJ}^{\langle -2,-2,0,0\rangle^+}
\right)
\right|_{*},
&\qquad& (m\ \mathrm{odd})
\\
0 &=
\partial_x^m \partial_t^n
\left.
\left(
g_{JTTJ}^{\langle 1,-2,-2,1\rangle^+}
    + g_{TTJJ}^{\langle -2,-2,1,1\rangle^+}
\right)
\right|_{*},
&\qquad& (m\ \mathrm{even})
\\
0 &=
\partial_x^m \partial_t^n
\left.
\left(
g_{JTTJ}^{\langle 1,-2,-2,1\rangle^+}
    + g_{TTJJ}^{\langle -2,-2,1,1\rangle^+}
\right)
\right|_{*},
&\qquad& (m\ \mathrm{odd})
\\
0 &=
\partial_x^m \partial_t^n
\left.
\left(
g_{JTTJ}^{\langle 0,-2,-1,-1\rangle^+}
    - g_{JTTJ}^{\langle 0,-1,-2,-1\rangle^+}
    + g_{TTJJ}^{\langle -1,-2,0,-1\rangle^+}
    - g_{TTJJ}^{\langle -2,-1,0,-1\rangle^+}
\right)
\right|_{*},
&\qquad& (m\ \mathrm{even})
\\
0 &=
\partial_x^m \partial_t^n
\left.
\left(
g_{JTTJ}^{\langle 0,-2,-1,-1\rangle^+}
    - g_{JTTJ}^{\langle 0,-1,-2,-1\rangle^+}
    + g_{TTJJ}^{\langle -1,-2,0,-1\rangle^+}
    - g_{TTJJ}^{\langle -2,-1,0,-1\rangle^+}
\right)
\right|_{*},
&\qquad& (m\ \mathrm{odd})
\\
0 &=
\partial_x^m \partial_t^n
\left.
\left(
g_{JTTJ}^{\langle -1,-2,0,-1\rangle^+}
    + g_{TTJJ}^{\langle 0,-2,-1,-1\rangle^+}
\right)
\right|_{*},
&\qquad& (m\ \mathrm{even})
\\
0 &=
\partial_x^m \partial_t^n
\left.
\left(
g_{JTTJ}^{\langle -1,-2,0,-1\rangle^+}
    + g_{TTJJ}^{\langle 0,-2,-1,-1\rangle^+}
\right)
\right|_{*},
&\qquad& (m\ \mathrm{odd})
\\
\intertext{\noindent\textbf{Line equations for $\langle JTTJ\rangle$ and $\langle TTJJ\rangle$}}
0 &=
\partial_x^m
\left.
\left(
g_{JTTJ}^{\langle 1,-2,-1,0\rangle^+}
    - g_{JTTJ}^{\langle 1,-1,-2,0\rangle^+}
    + g_{TTJJ}^{\langle -1,-2,1,0\rangle^+}
    - g_{TTJJ}^{\langle -2,-1,1,0\rangle^+}
\right)
\right|_{*},
&\qquad& (m\ \mathrm{even})
\\
0 &=
\partial_x^m
\left.
\left(
g_{JTTJ}^{\langle 1,-2,-1,0\rangle^+}
    - g_{JTTJ}^{\langle 1,-1,-2,0\rangle^+}
    + g_{TTJJ}^{\langle -1,-2,1,0\rangle^+}
    - g_{TTJJ}^{\langle -2,-1,1,0\rangle^+}
\right)
\right|_{*},
&\qquad& (m\ \mathrm{odd})
\\
0 &=
\partial_x^m
\left.
\left(
g_{JTTJ}^{\langle 1,-2,0,-1\rangle^+}
    + g_{JTTJ}^{\langle 1,0,-2,-1\rangle^+}
    + g_{TTJJ}^{\langle 0,-2,1,-1\rangle^+}
    + g_{TTJJ}^{\langle -2,0,1,-1\rangle^+}
\right)
\right|_{*},
&\qquad& (m\ \mathrm{even})
\\
0 &=
\partial_x^m
\left.
\left(
g_{JTTJ}^{\langle 1,-2,0,-1\rangle^+}
    + g_{JTTJ}^{\langle 1,0,-2,-1\rangle^+}
    + g_{TTJJ}^{\langle 0,-2,1,-1\rangle^+}
    + g_{TTJJ}^{\langle -2,0,1,-1\rangle^+}
\right)
\right|_{*},
&\qquad& (m\ \mathrm{odd})
\\
0 &=
\partial_x^m
\left.
\left(
g_{JTTJ}^{\langle 0,-2,0,0\rangle^+}
    + g_{TTJJ}^{\langle 0,-2,0,0\rangle^+}
\right)
\right|_{*},
&\qquad& (m\ \mathrm{even})
\\
0 &=
\partial_x^m
\left.
\left(
g_{JTTJ}^{\langle 0,-2,0,0\rangle^+}
    + g_{TTJJ}^{\langle 0,-2,0,0\rangle^+}
\right)
\right|_{*},
&\qquad& (m\ \mathrm{odd})
\\
0 &=
\partial_x^m
\left.
\left(
g_{JTTJ}^{\langle 1,-2,0,1\rangle^+}
    + g_{TTJJ}^{\langle 0,-2,1,1\rangle^+}
\right)
\right|_{*},
&\qquad& (m\ \mathrm{even})
\\
0 &=
\partial_x^m
\left.
\left(
g_{JTTJ}^{\langle 1,-2,0,1\rangle^+}
    + g_{TTJJ}^{\langle 0,-2,1,1\rangle^+}
\right)
\right|_{*},
&\qquad& (m\ \mathrm{odd})
\\
0 &=
\partial_x^m
\left.
\left(
g_{JTTJ}^{\langle 1,-2,1,0\rangle^+}
    - g_{JTTJ}^{\langle 1,1,-2,0\rangle^+}
    + g_{TTJJ}^{\langle 1,-2,1,0\rangle^+}
    - g_{TTJJ}^{\langle -2,1,1,0\rangle^+}
\right)
\right|_{*},
&\qquad& (m\ \mathrm{even})
\\
0 &=
\partial_x^m
\left.
\left(
g_{JTTJ}^{\langle 1,-2,1,0\rangle^+}
    - g_{JTTJ}^{\langle 1,1,-2,0\rangle^+}
    + g_{TTJJ}^{\langle 1,-2,1,0\rangle^+}
    - g_{TTJJ}^{\langle -2,1,1,0\rangle^+}
\right)
\right|_{*},
&\qquad& (m\ \mathrm{odd})
\\
\intertext{\noindent\textbf{Point equations for $\langle JTTJ\rangle$ and $\langle TTJJ\rangle$}}
0 &=
\left.
\left(
g_{JTTJ}^{\langle -1,-2,2,-1\rangle^+}
    + g_{TTJJ}^{\langle 2,-2,-1,-1\rangle^+}
\right)
\right|_{*},
&\qquad&
\\
0 &=
\left.
\left(
g_{JTTJ}^{\langle 0,-2,1,-1\rangle^+}
    - g_{JTTJ}^{\langle 0,1,-2,-1\rangle^+}
    + g_{TTJJ}^{\langle 1,-2,0,-1\rangle^+}
    - g_{TTJJ}^{\langle -2,1,0,-1\rangle^+}
\right)
\right|_{*},
&\qquad&
\\
0 &=
\left.
\left(
g_{JTTJ}^{\langle 1,-2,2,-1\rangle^+}
    + g_{JTTJ}^{\langle 1,2,-2,-1\rangle^+}
    + g_{TTJJ}^{\langle 2,-2,1,-1\rangle^+}
    + g_{TTJJ}^{\langle -2,2,1,-1\rangle^+}
\right)
\right|_{*}.
&\qquad&
\end{alignat*}

}

\section{Three point structures}\label{ThreePointStructures}

\subsection{$JJ\mathcal{O}$}
\begin{center}
	\begin{tabular}{|c|c|c|}
		\hline
		Parity                  & $j_3$                            & Structure                              \\
		\hline
		\multirow{3}{2em}{Even} & 0                                & \texttt{JJOScalarParityEven}           \\
		\cline{2-3}
		                        & \multirow{2}{4em}{$\geq 2$ even} & \texttt{JJOGenericEvenSpinParityEven1} \\
		                        &                                  & \texttt{JJOGenericEvenSpinParityEven2} \\
		\hline
		\multirow{3}{2em}{Odd}  & 0                                & \texttt{JJOScalarParityOdd}            \\
		\cline{2-3}
		                        & $\geq 2$ even                    & \texttt{JJOGenericEvenSpinParityOdd}   \\
		\cline{2-3}
		                        & $\geq 3$ odd                     & \texttt{JJOGenericOddSpinParityOdd}    \\
		\hline
	\end{tabular}
\end{center}

\subsection{$JT\mathcal{O}$}
\begin{center}
	\begin{tabular}{|c|c|c|}
		\hline
		Parity                  & $j_3$                       & Structure                          \\
		\hline
		\multirow{4}{2em}{Even} & 1                           & \texttt{JTOSpin1ParityEven}        \\
		\cline{2-3}
		                        & 2                           & \texttt{JTOSpin2ParityEven}        \\
		\cline{2-3}
		                        & \multirow{2}{2em}{$\geq 3$} & \texttt{JTOGenericSpinParityEven1} \\
		                        &                             & \texttt{JTOGenericSpinParityEven2} \\
		\hline
		\multirow{4}{2em}{Odd}  & 1                           & \texttt{JTOSpin1ParityOdd}         \\
		\cline{2-3}
		                        & 2                           & \texttt{JTOSpin2ParityOdd}         \\
		\cline{2-3}
		                        & \multirow{2}{2em}{$\geq 3$} & \texttt{JTOGenericSpinParityOdd1}  \\
		                        &                             & \texttt{JTOGenericSpinParityOdd2}  \\
		\hline
	\end{tabular}
\end{center}

\subsection{$TJ\mathcal{O}$}
\begin{center}
	\begin{tabular}{|c|c|c|}
		\hline
		Parity                  & $j_3$                       & Structure                          \\
		\hline
		\multirow{4}{2em}{Even} & 1                           & \texttt{TJOSpin1ParityEven}        \\
		\cline{2-3}
		                        & 2                           & \texttt{TJOSpin2ParityEven}        \\
		\cline{2-3}
		                        & \multirow{2}{2em}{$\geq 3$} & \texttt{TJOGenericSpinParityEven1} \\
		                        &                             & \texttt{TJOGenericSpinParityEven2} \\
		\hline
		\multirow{4}{2em}{Odd}  & 1                           & \texttt{TJOSpin1ParityOdd}         \\
		\cline{2-3}
		                        & 2                           & \texttt{TJOSpin2ParityOdd}         \\
		\cline{2-3}
		                        & \multirow{2}{2em}{$\geq 3$} & \texttt{TJOGenericSpinParityOdd1}  \\
		                        &                             & \texttt{TJOGenericSpinParityOdd2}  \\
		\hline
	\end{tabular}
\end{center}

\subsection{$JJT$}
\begin{center}
	\begin{tabular}{|c|c|c|}
		\hline
		OPE coefficient                          & ordered triple $\langle \mathcal{O}_i \mathcal{O}_j \mathcal{O}_k \rangle$ & three-point structure                              \\
		\hline
		\multirow{3}{4em}{$\lambda_B+\lambda_F$} & $\langle TJJ \rangle$                                                      & \texttt{TJJDiracFermion}+\texttt{TJJComplexScalar} \\
		                                         & $\langle JTJ \rangle $                                                     & \texttt{JTJDiracFermion}+\texttt{JTJComplexScalar} \\
		                                         & $\langle JJT \rangle $                                                     & \texttt{JJTDiracFermion}+\texttt{JJTComplexScalar} \\
		\hline
		\multirow{3}{4em}{$\lambda'_{JJT}$}      & $\langle TJJ \rangle$                                                      & \texttt{TJJDiracFermion}-\texttt{TJJComplexScalar} \\
		                                         & $\langle JTJ \rangle$                                                      & \texttt{JTJDiracFermion}-\texttt{JTJComplexScalar} \\
		                                         & $\langle JJT \rangle$                                                      & \texttt{JJTDiracFermion}-\texttt{JJTComplexScalar} \\
		\hline
	\end{tabular}
\end{center}

\subsection{$TTT$}
\begin{center}
	\begin{tabular}{|c|c|c|}
		\hline
		OPE coefficient                  & ordered triple $\langle \mathcal{O}_i \mathcal{O}_j \mathcal{O}_k \rangle$ & three-point structure  \\
		\hline
		\multirow{1}{4em}{$\lambda_B$}   & $\langle TTT \rangle$                                                      & \texttt{StressTensorB} \\
		\hline
		\multirow{1}{4em}{$\lambda_{F}$} & $\langle TTT \rangle$                                                      & \texttt{StressTensorF} \\
		\hline
	\end{tabular}
\end{center}

\bibliographystyle{JHEP}
\bibliography{refs}
\end{document}